\documentclass[cameraready]{Interspeech}

\title{Evaluating Parkinson’s Disease Detection in Anonymized Speech: A Performance and Acoustic Analysis}

\author[affiliation={1,2}]{Carlos}{Franzreb}
\author[affiliation={2}]{Francisco}{Teixeira}
\author[affiliation={3,4}]{Ben}{Luks}
\author[affiliation={3}]{Sebastian}{Möller}
\author[affiliation={2,4}]{Alberto}{Abad}

\address{
    $^1$ German Research Center for Artificial Intelligence, Germany \\
    $^2$ INESC-ID, Portugal \\
    $^3$ Technical University of Berlin, Germany \\
    $^4$ Instituto Superior Técnico, University of Lisbon, Portugal
}

\email{carlos.franzreb@dfki.de}


\usepackage{comment}
\usepackage{multirow}
\usepackage[subtle]{savetrees}

\begin{document}

\maketitle

\begin{abstract}
Automatic detection of Parkinson's disease (PD) from speech is a promising non-invasive diagnostic tool, but it raises significant privacy concerns.
Speaker anonymization mitigates these risks, but it may suppress the pathological information necessary for PD detection.
We assess the trade-off between privacy and PD detection for two anonymizers (STT-TTS and kNN-VC) using two Spanish datasets.
STT-TTS provides better privacy but severely degrades PD detection by eradicating prosodic information.
kNN-VC preserves macro-prosodic features such as duration and F0 contours, achieving F1 scores only 3-7\% lower than original baselines, demonstrating that privacy-preserving PD detection is viable when using appropriate anonymization.
Finally, an acoustic distortion analysis characterizes specific weaknesses in kNN-VC, offering insights for designing anonymizers that better preserve PD information.
\end{abstract}

\section{Introduction}
Parkinson's disease (PD) is a neurodegenerative disease of the central nervous system \cite{parkinson_essay_2002}, whose prevalence is rapidly increasing \cite{zhu_temporal_2024}.
It affects both movement, with motor symptoms such as tremors, and speech production \cite{moustafa_motor_2016}.
Stuttering, slurring, and an uncontrollable increase in speaking rate are common in people with PD.
Previous research has found that PD patients exhibit articulatory difficulties~\cite{ackermann_articulatory_1991}.
A larger study~\cite{skodda_progression_2013} reports differences in articulation, voice, tempo, and fluency between PD patients and a healthy control group.

Treatment of PD is more effective when detected early, but the initial lack of symptoms makes its detection difficult.
There is no definitive laboratory test for identifying PD, and accurate identification by physicians requires deep expertise \cite{post_unified_2005}.
These difficulties have spurred research on automatic PD detection \cite{pahuja_comparative_2021}, including detection from speech \cite{gimeno-gomez_unveiling_2025, quan_deep_2021, la_quatra_bilingual_2025}, opening an avenue for less invasive and more cost-effective automatic PD detection methods.

Besides identifiers relevant for PD detection, speech carries extensive biometric information that raises privacy concerns \cite{nautsch_gdpr_2019}.
Collecting data for training deep learning models is difficult in the face of regulations, and users may refuse to use services that capture their identity \cite{franzreb_towards_2024}.
Speaker anonymization \cite{jin_voice_2009} offers a solution to this problem, concealing the speaker's identity while preserving the utility of the speech.
How utility is characterized depends on the use case \cite{meyer_use_2025}; in the case of PD speech, anonymization can either conceal pathologies to protect the speaker's identity, or preserve them to enable their detection while concealing the remaining sensitive information.
The latter is more challenging, as PD information and speaker identity are entangled \cite{champion_are_2022}.

To quantify how much PD information is preserved, we leverage an existing detection algorithm \cite{gimeno-gomez_unveiling_2025}.
We consider two speech datasets comprising PD patients (PC-GITA \cite{orozco-arroyave_new_2014} and Neurovoz \cite{mendes-laureano_neurovoz_2024}) and two anonymization approaches: speech-to-text-to-speech (STT-TTS) and kNN-VC \cite{baas_voice_2023}.
We also compare the acoustic features of original and anonymized speech, to better understand which PD information is lost, and estimate the privacy gain provided to PD speakers.
The outcome of this analysis may inform the design of future anonymizers that better preserve PD information.
All the code necessary to reproduce our experiments is available online\footnote{\url{https://github.com/carlosfranzreb/spane}}.

\subsection{Related work}

Several studies leverage anonymization to conceal pathological traits, motivated by privacy, stigma reduction, or fair access to speech-driven technologies.
Voice-conversion-based anonymization has been compared with STT–TTS for stuttered speech \cite{hintz_anonymization_2023}, showing that text-based resynthesis can effectively remove stuttering characteristics altogether, resulting in strong privacy while masking the pathology.
More broadly, previous studies have argued that strong anonymization or privacy-preserving transformations can suppress clinically relevant acoustic abnormalities, potentially masking neurological or psychiatric conditions~\cite{diaz-asper_navigating_2025}.

In contrast, other studies have investigated whether pathological characteristics remain identifiable after anonymization, primarily to assess the utility of anonymized speech for clinical research.
A large-scale evaluation across multiple speech disorders \cite{tayebi_arasteh_addressing_2024} showed that, although anonymization degrades speaker verification performance, downstream pathology-related tasks, such as disorder classification and intelligibility assessment, are largely preserved for many conditions, including dysarthria and dysphonia.
Furthermore, a subjective evaluation showed that pathology-relevant acoustic attributes such as hoarseness and tremor can survive anonymization when explicitly modeled, allowing anonymized speech to remain clinically informative~\cite{ghosh_anonymising_2024}.
These studies suggest that anonymization does not necessarily eliminate pathological markers and that, with careful design, anonymized speech can still support pathology identification and analysis.
However, both studies evaluate privacy with speaker recognizers trained on original speech, which has been shown to overestimate privacy \cite{champion_anonymizing_2023}.
Furthermore, they do not investigate how anonymization acoustically alters the utterances

While detecting PD from anonymized speech is promising, it has its drawbacks, such as the potential loss of relevant clinical information.
There are alternative approaches that do not degrade the acoustic signal, such as cryptographic tools \cite{teixeira2019privacy}, which process the signal in a confidential manner, or privacy-aware feature extraction methods \cite{ravi2024enhancing}, which run on the user's device and only share clinically relevant information.
However, the former are computationally expensive and require complex configurations to ensure privacy \cite{teixeira2024privacy}, while the latter often lack interpretability and are task-specific.
In contrast, anonymized speech can be processed in the same way as original speech, and thus provides flexibility and allows for transparent analysis.

\section{Methodology}

In this work, we first attempt to evaluate popular anonymizers on PD detection.
We propose PD detection as a novel utility evaluation for anonymizers, to assess whether PD can be identified from anonymized speech, and investigate how PD affects the speaker's privacy.
We consider two datasets: PC-GITA \cite{orozco-arroyave_new_2014}, in Colombian Spanish, and Neurovoz \cite{mendes-laureano_neurovoz_2024}, in Castilian Spanish.
We focus on two tasks: \textit{sentences}, where participants utter a fixed set of sentences, and \textit{monologue}, where participants speak freely for approximately half a minute, either talking about their daily routine (PC-GITA) or describing an image (Neurovoz).
For brevity, we refer to PC-GITA as GT and to Neurovoz as NV, appending the first letter of the task (e.g. PC-GITA \textit{sentences} is abbreviated as GT-S).
GT-S is designed to assess the prosody of the speakers, who read phonetically balanced sentences aloud.
NV-S comprises popular sayings, and the speakers listen to them instead of reading them, to account for vision impairments.

We consider two anonymization approaches: speech-to-text-to-speech (STT-TTS) and kNN-VC \cite{baas_voice_2023}.
STT-TTS consists of the concatenation of Whisper-large \cite{radford_robust_2022} ASR and Kokoro\footnote{\url{https://github.com/hexgrad/kokoro}} TTS, resulting in the removal of all speaker information by transcribing the input speech before synthesis \cite{franzreb_content_2026}.
kNN-VC \cite{baas_voice_2023} offers worse privacy but higher fidelity to the source speech, as it preserves prosodic information \cite{gengembre_disentangling_2024, franzreb_private_2025}.
It is an any-to-any voice converter: features extracted with WavLM \cite{chen_wavlm_2022} are replaced with the average of the four nearest target features according to cosine similarity and then synthesized with a neural vocoder (HiFiGAN \cite{kong_hifi-gan_2020}).

kNN-VC requires speech from a target speaker to perform the conversion.
We use a crowd-sourced Colombian (CC) dataset \cite{guevara_2020_crowdsourcing}, which comprises 17 speakers, each recorded for 27 minutes on average.
Preliminary experiments with the PD detector trained on original speech showed that same-gender selection with the CC dataset performs better than cross-gender or an unconstrained selection.
Using PC-GITA target speakers did not perform better, even when the speaker's condition (HC or PD) was preserved.
We therefore use same-gender selection with the CC dataset for kNN-VC, and also for STT-TTS with its two available target speakers (one per gender).
A target speaker is selected randomly for each utterance.

\section{Utility evaluation}

The PD evaluation we use is based on an automatic PD detector (PDD) \cite{gimeno-gomez_unveiling_2025}, which trains a classifier using Wav2vec2 features \cite{baevski_wav2vec_2020}.
The PDD is trained and evaluated for each task independently with a 5-fold cross-validation, and each experiment is repeated 5 times with different seeds.
Given the scarcity of data, the PDD does not generalize well to other datasets or tasks, but shows strong performance for several datasets in various languages.
We therefore also consider each task independently, but we consider two evaluation scenarios:

\begin{enumerate}
    \item \textbf{PDD-O:} the PDD is trained on original speech.
    \item \textbf{PDD-A:} the PDD is trained on speech anonymized by the same anonymizer under evaluation.
\end{enumerate}

PDD-O represents the simplest use case: someone wants to pass their speech to the PDD without the associated privacy risk.
PDD-A requires the service provider to train the PDD with anonymized data and is also restricted to work with that particular anonymizer.
For the purpose of this study, PDD-O assesses the generalizability of the PDD and the anonymizer, while PDD-A quantifies how much PD information is present in the anonymized speech.

\begin{table}[th]
\centering
\caption{PDD F1-scores for different gender-preserving anonymizers, datasets and tasks. O and A refer to the training data of the PDD (original or anonymized). The PDD-A results of original data are calculated with kNN-VC's PDD-A. WERs are computed with Whisper-large only for the sentences tasks (S), as there is no ground truth for the monologue tasks (M).}
\label{tab:pdd_results}
\scalebox{0.9}{
\begin{tabular}{llccc}
\toprule
\textbf{Dataset} & \textbf{Anon.} & \textbf{PDD-O} & \textbf{PDD-A} & \textbf{WER} \\
\midrule
\multirow{4}{*}{GT-S}
    & None       & $79\pm1$ & $78\pm1$ & $13$ \\
    & kNN-VC$_r$ & $47\pm1$ & $80\pm1$ & $18$ \\
    & kNN-VC     & $54\pm1$ & $76\pm0$ & $21$ \\
    & STT-TTS    & $49\pm0$ & $53\pm1$ & $12$ \\
\midrule
\multirow{4}{*}{GT-M} 
    & None       & $80\pm3$ & $64\pm4$ & --- \\
    & kNN-VC$_r$ & $49\pm3$ & $80\pm1$ & --- \\
    & kNN-VC     & $55\pm4$ & $74\pm1$ & --- \\
    & STT-TTS    & $35\pm2$ & $49\pm1$ & --- \\
\midrule

\multirow{4}{*}{NV-S} 
    & None       & $86\pm0$ & $81\pm1$ & $8$ \\
    & kNN-VC$_r$ & $81\pm2$ & $86\pm1$ & $12$ \\
    & kNN-VC     & $70\pm2$ & $82\pm1$ & $17$ \\
    & STT-TTS    & $35\pm0$ & $64\pm0$ & $12$ \\
\midrule
\multirow{4}{*}{NV-M} 
    & None       & $85\pm3$ & $56\pm8$ & --- \\
    & kNN-VC$_r$ & $76\pm2$ & $82\pm2$ & --- \\
    & kNN-VC     & $70\pm9$ & $78\pm1$ & --- \\
    & STT-TTS    & $54\pm5$ & $59\pm1$ & --- \\

\bottomrule
\end{tabular}
}
\vspace{-0.3cm}
\end{table}

\subsection{PDD results}

Table \ref{tab:pdd_results} displays the PDD F1 scores for both datasets and tasks.
Word error rates (WERs) are computed with Whisper-large to ensure that the anonymizer's output is intelligible, which is the case for both anonymizers and datasets.
On average, the intelligibility estimated for PD speakers is up to three times higher than that for the healthy control group (e.g. WER$_{HC} = 10$ and WER$_{PD} = 31$ for the original GT-S).
The baseline PDD-O (i.e., the PDD trained and evaluated with original speech) shows higher F1 scores on Neurovoz than PC-GITA for both tasks, but the scores for each dataset are consistent between tasks.
These scores are equivalent to those reported by previous work \cite{gimeno-gomez_unveiling_2025}.

F1 scores with original speech on the PDD-A column refer to an evaluation using original speech and the PDD trained on kNN-VC.
Interestingly, it performs well for the \textit{sentences} task, implying that a PDD trained on anonymized speech could be used to detect PD from original speech.
For the \textit{monologue} task, the performance degradation is greater.

STT-TTS proves ineffective as an anonymizer for PD detection, as the prosodic and acoustic information required is removed in the transcription \cite{hintz_anonymization_2023}.
Notably, because Whisper-large does not preserve stutters and hesitations, so only transcription errors can be used to discriminate HC from PD speakers.
As expected, STT-TTS achieves low PDD-A F1 scores ($< 54\%$) for both PC-GITA tasks.
For Neurovoz the PDD-A scores are higher (59\% for NV-M, 64\% for NV-S).
The higher performance could arise from the increased training data (1,696 instead of 1,000), which allows the PDD-A to pinpoint re-occurring transcription errors.
Neurovoz also includes two sentences with WER=0 for all HC speakers anonymized with STT-TTS, which does not occur in PC-GITA.

kNN-VC's PDD-O scores are considerably lower than those for the original utterances: 25\% lower for PC-GITA and 15\% for Neurovoz.
The F1 scores for Neurovoz are still high (70\% each).
The PDD-A F1 scores show little degradation for the \textit{sentences} task compared to the baseline PDD-O: 3\% for PC-GITA, and 4\% for Neurovoz.
The \textit{monologue} task proves to be more challenging for anonymizers, with larger differences to the baseline: 6\% for PC-GITA, and 7\% for Neurovoz.
The \textit{sentences} task may be easier because speakers utter the same sentences, which leads to all speakers making similar prosodic and pronunciation errors.
This hypothesis also explains the generalizability of kNN-VC's PDD-A to original speech.

\subsection{Analyzing the impact of the conversion step in kNN-VC}

Given the similar nature of the feature extractors of kNN-VC and the PDD (WavLM and Wav2vec2), the loss of PD information likely takes place in the last two steps: the conversion step or the vocoder.
To investigate the loss incurred by each step, we run a resynthesis version of kNN-VC (called kNN-VC$_r$ in Table \ref{tab:pdd_results}) where the WavLM features are passed to the vocoder, bypassing the conversion step.
The PDD-A results suggest that the loss of PD information occurs in the conversion step: kNN-VC$_r$ achieves the same F1 score as the original baseline for all tasks except NV-M, where it still outperforms kNN-VC by 4\%.

The PDD-O results differ depending on the dataset.
For Neurovoz, the scores of kNN-VC$_r$ are substantially better than those of kNN-VC, with differences of 11\% for NV-S and 6\% for NV-M, although the standard deviation for the latter is larger (9\%).
The different accents between source and target speakers in kNN-VC (Castilian and Colombian) likely contribute to its lower scores.
However, for PC-GITA, kNN-VC$_r$ performs worse than kNN-VC for both tasks.
As for Neurovoz, the difference is only large for the \textit{sentences} task (7\%).
These results fit our preliminary experiments with target speakers, where kNN-VC$_r$'s performance falls between kNN-VC with target speakers from the CC dataset and kNN-VC with PC-GITA targets.
The performance drop may stem from an acoustic mismatch between PC-GITA and Librispeech, the vocoder's training data.
Converting the PC-GITA utterances to CC target speakers potentially alleviates the mismatch.

\section{Privacy evaluation}

\begin{table}[]
    \centering
    \caption{Intra-EERs as percentages of the privacy evaluation for original speech and the two anonymizers. The English datasets are evaluated before the MLS-ES fine-tuning.}
    \scalebox{0.9}{
    \begin{tabular}{lcccccc}
        \toprule
        & \multicolumn{6}{c}{Anonymizer} \\
        \cmidrule{2-7}
        &
        \multicolumn{2}{c}{\textbf{None}} &
        \multicolumn{2}{c}{\textbf{kNN-VC}} &
        \multicolumn{2}{c}{\textbf{STT-TTS}} \\
        \cmidrule(lr){2-3}
        \cmidrule(lr){4-5}
        \cmidrule(lr){6-7}
        \textbf{Dataset} &
        HC & PD & HC & PD & HC & PD \\
        \midrule
            LS & $0.5$ & --- & $6.9$ & --- & $35.6$ & --- \\
            EdAcc & $3.0$ & --- & $17.9$ & --- & $43.3$ & --- \\
        \midrule
            MLS-ES & $0.2$ & --- & $2.4$ & --- & $26.0$ & --- \\
        \midrule
            GT-S & $1.0$ & $1.2$ & $20.8$ & $21.2$ & $36.4$  & $38.4$ \\
            GT-M & $0.9$ & $1.5$ & $11.8$ & $10.2$ & $28.3$  & $30.3$ \\
        \midrule
            NV-S & $1.4$ & $1.6$ & $23.9$ & $24.7$ & $55.8$ & $58.5$ \\
            NV-M & $1.8$ & $0.0$ & $13.4$ & $16.6$ & $49.3$ & $51.2$ \\
        \bottomrule
    \end{tabular}}
    \label{tab:privacy_results}
    \vspace{-0.2cm}
\end{table}

We now investigate how PD affects privacy estimates.
PD speakers may be easier to identify, as disfluencies present a new risk of identity leakage, but only if the speaker recognizer can leverage them.
We use a privacy evaluation \cite{franzreb_optimizing_2025} based on the VoicePrivacy Challenge 2024 protocol \cite{tomashenko2024voiceprivacy}, which simulates an attack on anonymized speech with a speaker recognizer (ECAPA-TDNN \cite{dawalatabad_ecapa-tdnn_2021}).
Each anonymizer has a dedicated speaker recognizer trained on its anonymized data.
Given the small size of the two PD datasets, we first train the recognizers with Librispeech \textit{train-clean-360} and then fine-tune them with the Spanish set of Multilingual Librispeech (MLS-ES) \cite{pratap20_interspeech}.
Training directly with MLS-ES is not optimal due to its low number of speakers.
MLS-ES is heavily imbalanced; we address this by only training on up to 1,000 utterances per speaker.

Following a study on how to configure evaluation datasets \cite{franzreb_optimizing_2025}, we use 20 trial and 20 enrollment utterances from the test set of MLS-ES.
For the PD datasets, we split all available utterances randomly and evenly into the trials and enrollments, which for the \textit{monologue} tasks means splitting the single long utterance available per speaker in half.
The  equal error rates (EERs) for all datasets and anonymizers are depicted in Table \ref{tab:privacy_results}.
For the PD datasets, we evaluate each group (HC and PD) separately with the intra-EER \cite{franzreb_content_2026}, which assumes that the attacker knows which group the speaker belongs to, and assesses how identifiable each speaker is compared to the other members of its group.

The results with original speech validate the privacy evaluation: the recognizer can verify almost all pairs of MLS-ES utterances.
The increased privacy provided by the two anonymizers is also apparent in the MLS-ES results.
The privacy estimated for STT-TTS is suboptimal (EER$<50\%$) because the distinct content uttered by each speaker in MLS-ES leads to phonetic differences, which can be captured by the speaker recognizer \cite{franzreb_content_2026}.

The results for the two PD datasets show similar privacy improvements from the anonymizers.
For kNN-VC, privacy estimates are consistently higher than for original speech.
The privacy estimates for STT-TTS are by far the largest, achieving near-perfect anonymization for the Neurovoz tasks (EER$\approx50\%$).
The evaluation is able to identify speakers anonymized with STT-TTS in both PC-GITA tasks, suggesting that some speaker information remains in the transcriptions.
Interestingly, these findings contrast with the PDD results, where PD information remained in some Neurovoz transcriptions.

PD speakers achieve higher privacy estimates in all cases except for original NV-M and kNN-VC's GT-M.
Although the differences are small, the speaker recognizer apparently struggles to characterize PD speakers, which can be explained by the lack of pathological speech in MLS-ES.
Considering all three anonymization scenarios, the Neurovoz EERs are larger overall than those of PC-GITA, suggesting that PC-GITA speakers are more distinct.
This result also explains why PC-GITA's PDD F1 scores were lower than those of Neurovoz, as the greater speaker variability makes finding speaker-agnostic features more challenging.
Regarding the two tasks, the EERs for \textit{sentences} are always higher than those for \textit{monologue}; the speaker recognizer performs better for longer utterances.

\begin{figure*}
    \centering
    \includegraphics[width=0.9\textwidth]{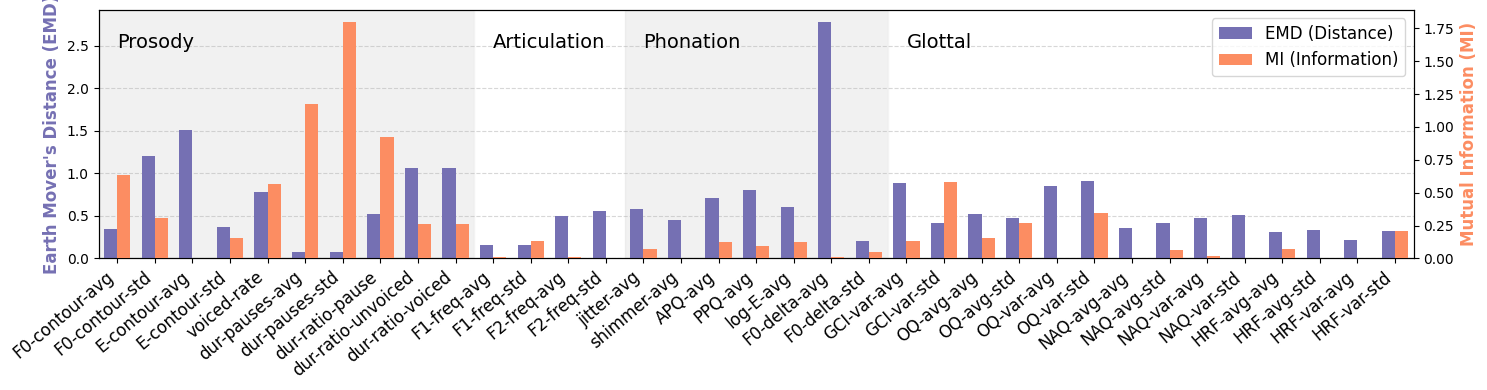}
    \vspace{-0.2cm}
    \caption{EMD and MI per feature, for the monologue task of PC-GITA.}
    \label{fig:metrics_monologue}
    \vspace{-0.4cm}
\end{figure*}

\section{Distortion analysis}

To better understand which PD information is retained by kNN-VC, we compare the acoustic features of original and anonymized samples for GT-M, whose features are more reliable due to their longer duration.
We consider a subset of 35 DisVoice features \cite{vasquez-correa_towards_2018} which is particularly relevant for PD detection \cite{gimeno-gomez_unveiling_2025}.
They include prosodic, glottal, phonation, and articulatory features.
A standard scaler was fit to the original features and used to transform both the original and anonymized features.
This normalization allows for a direct comparison between features.

To quantify the magnitude of acoustic distortion introduced by kNN-VC, we compare the original and anonymized utterances using two complementary metrics: Earth Mover's Distance (EMD) \cite{rubner1998metric} and Mutual Information (MI) \cite{cover_elements_2001}.
EMD quantifies the cost of transforming the original distribution into the anonymized distribution.
MI quantifies the amount of non-linear information shared between the original and anonymized features.
In our analysis, MI reveals whether the underlying pathological severity survived the anonymization process, even if the absolute acoustic values were shifted.
Regarding our PDD evaluations, MI is a proxy for PDD-A, describing whether the required information is available for performing PDD.
On the other hand, EMD is a proxy for PDD-O, describing whether anonymized utterances can be evaluated with the original PDD.

The EMD and MI values are plotted in Figure \ref{fig:metrics_monologue}.
The most severe distortion in the dataset occurred within the phonation category: the avg. of the F0 derivative exhibits the largest EMD ($2.78$) alongside a near-zero MI ($0.01$).
The actual values are lower for the anonymized speech, meaning that the pitch contour is smoother, presumably without the vocal tremors and pitch breaks that occur in PD speech, as these are not present in healthy target speech.
Articulation and glottal features exhibit consistently low MI scores (mostly $<0.10$).
The anonymized articulation and glottal features contain information about the healthy target's vocal tract, not the PD patient's, rendering these features uninformative for post-anonymization classification.
The only exception is the standard deviation of the GCI variability ($MI=0.58$), which is a robust measure for jitter.
Its large EMD value ($0.25$), as well as the poor preservation of the avg. GCI variability, suggest that the standard deviations of the GCI variability are altered by kNN-VC, but that the differences among utterances remain.

Prosody is the category that retains the most PD information, though mainly within macro-temporal boundaries.
The avg. and std. of the duration pauses achieve the highest MI values (1.18 and 1.80) and low EMD values ($<0.08$), meaning that they are not altered by kNN-VC and that PDD-O should also be able to capture them.
Overall, the frame-by-frame mapping of kNN-VC successfully maintains the broader durations of the source speaker.
Conversely, micro-prosodic elements are distorted.
The unvoiced duration ratio shifts considerably (EMD=1.06), supporting the hypothesis that kNN-VC suppresses pathological mid-syllable micro-pauses and breathy asthenia.
The avg. energy contour is also completely altered by kNN-VC (EMD=1.51, MI=0.00), with the anonymized speech having higher energy on average, presumably due to the healthy target speech.
Finally, the average F0 contour is also moderately preserved (EMD=0.35, MI=0.63), as were the GCI fluctuations.
Pathological traits such as monopitch or irregular breath groups can be used in part to detect PD from anonymized speech.
Both these features can be leveraged by PDD-A to detect PD from anonymized speakers, but likely not by PDD-O.

\section{Discussion}

In this work, we performed a thorough analysis of the impact and potential of anonymization in the detection of Parkinson's disease (PD) from speech.
%
%
Our results showed that detecting PD from speech anonymized with kNN-VC with PDD-O is not viable;
the PDD has to be trained on anonymized speech for the evaluation to adequately reflect the PD information preserved by the anonymizer.
Interestingly, the PDD-A trained on kNN-VC speech is also effective when evaluating original speech for the \textit{sentences} task, as the constrained phonetics of the task allow for a more targeted identification of PD information.
This means that PDDs can be trained on anonymized speech, which is easier to collect, with minimal performance degradation.
For this task, the PDDs are not required to learn features that generalize to other sentences, causing prosodic features to become more relevant, as each sentence is expected to follow a similar prosodic pattern.
Nevertheless, training PDDs with both original and anonymized utterances may improve their robustness and make them useful for both kinds of speech.
On the other hand, STT-TTS did not preserve PD information, agreeing with previous work \cite{hintz_anonymization_2023}.

The distortion analysis emphasized that kNN-VC fails to preserve glottal, articulatory, and phonation features, with the exception of the GCI variability, which is nevertheless shifted.
Only prosodic features are well preserved, particularly durations and the avg. F0 contour.
Still, this information is enough to achieve PDD-A F1 scores that are only 3\% to 7\% lower than the original baseline.
This result implies that articulatory and phonation features are not crucial for PD detection, as they are not well preserved by kNN-VC.

The resynthesis experiment showed that this loss of PD information is almost entirely due to the conversion step, where the source speech is replaced by healthy target speech.
Conversion could be improved by creating a pool of target speakers that comprises acoustic characteristics like micro-tremors and abrupt interruptions are required to represent PD information.
The kNN algorithm may also fail to preserve the relevant acoustic information.
WavLM features are dominated by their phonetic content \cite{sicherman_analysing_2023}, and it is unclear whether, considering target features representing the same phone, kNN selects those that preserve PD information.
It is also unclear how much PD information WavLM captures, and the same applies to the PDD's Wav2vec2; it is enough to achieve SOTA PD detection, but it may have some shortcomings, like the acoustic features we found to be misrepresented in the anonymized utterances.


Regarding the privacy evaluation, the speaker recognizer failed to take full advantage of the PD information to identify speakers.
Including pathological speech in the recognizer's training data should improve its ability to characterize PD speakers.

\section{Conclusion}

STT-TTS is estimated to provide near-perfect privacy for Neurovoz, and much higher privacy than kNN-VC for PC-GITA.
This anonymizer is well suited for use cases where the pathology should be anonymized along with the speaker.
For use cases that aim to detect PD while preserving the speaker's privacy, kNN-VC offers a good balance between PDD accuracy and privacy.
Overall, this study adds to the evidence that anonymization is a strong contender as a path forward to help ensure privacy in the detection of PD.
The conducted acoustic analysis may be replicated in other speech-affecting diseases, potentially extending the use cases of anonymization for speech-affecting disease detection.

\section{Acknowledgements}
This work was supported in part by the Federal Ministry of Education and Research, Germany
(BMBF 16KIS2048), the Portuguese national funds through Fundação para a Ciência e a Tecnologia, I.P. (FCT) under projects UID/50021/2025 (DOI: \url{https://doi.org/10.54499/UID/50021/2025}) and UID/PRR/50021/2025 (DOI: \url{https://doi.org/10.54499/UID/PRR/50021/2025}) and by the Portuguese Recovery and Resilience Plan and NextGenerationEU European Union funds under project C644865762-00000008 (Accelerat.AI).
\bibliographystyle{IEEEtran}
\bibliography{references}

@article{gimeno-gomez_unveiling_2025,
    title = {Unveiling {Interpretability} in {Self}-{Supervised} {Speech} {Representations} for {Parkinson}'s {Diagnosis}},
    journal = {IEEE Journal of Selected Topics in Signal Processing},
    author = {Gimeno-Gómez, David and Botelho, Catarina and Pompili, Anna and Abad, Alberto and Martínez-Hinarejos, Carlos-D.},
    year = {2025},
    pages = {1--14},
}

@article{moustafa_motor_2016,
    title = {Motor symptoms in {Parkinson}'s disease: {A} unified framework},
    volume = {68},
    journal = {Neuroscience \& Biobehavioral Reviews},
    author = {Moustafa, A. A. and Chakravarthy, S. and Phillips, J. R. and Gupta, A. and Keri, S. and Polner, B. and Frank, M. J. and Jahanshahi, M.},
    year = {2016},
    pages = {727--740},
}

@article{zhu_temporal_2024,
    title = {Temporal trends in the prevalence of {Parkinson}'s disease from 1980 to 2023: a systematic review and meta-analysis},
    volume = {5},
    number = {7},
    journal = {The Lancet. Healthy Longevity},
    author = {Zhu, Jinqiao and Cui, Yusha and Zhang, Junjiao and Yan, Rui and Su, Dongning and Zhao, Dong and Wang, Anxin and Feng, Tao},
    year = {2024},
    pages = {464--479},
}

@article{parkinson_essay_2002,
    title = {An essay on the shaking palsy. 1817},
    volume = {14},
    number = {2},
    journal = {The Journal of Neuropsychiatry and Clinical Neurosciences},
    author = {Parkinson, James},
    year = {2002},
    pmid = {11983801},
    pages = {223--236},
}

@article{post_unified_2005,
    title = {Unified {Parkinson}'s disease rating scale motor examination: are ratings of nurses, residents in neurology, and movement disorders specialists interchangeable?},
    volume = {20},
    number = {12},
    journal = {Movement Disorders: Official Journal of the Movement Disorder Society},
    author = {Post, Bart and Merkus, Maruschka P. and de Bie, Rob M. A. and de Haan, Rob J. and Speelman, Johannes D.},
    year = {2005},
    pmid = {16116612},
    pages = {1577--1584},
}

@article{quan_deep_2021,
    title = {A {Deep} {Learning} {Based} {Method} for {Parkinson}’s {Disease} {Detection} {Using} {Dynamic} {Features} of {Speech}},
    volume = {9},
    journal = {IEEE Access},
    author = {Quan, Changqin and Ren, Kang and Luo, Zhiwei},
    year = {2021},
    pages = {10239--10252},
}

@inproceedings{la_quatra_bilingual_2025,
    title = {Bilingual {Dual}-{Head} {Deep} {Model} for {Parkinson}’s {Disease} {Detection} from {Speech}},
    booktitle = {ICASSP},
    author = {La Quatra, Moreno and Orozco-Arroyave, Juan Rafael and Siniscalchi, Marco Sabato},
    year = {2025},
    pages = {1--5},
}

@article{pahuja_comparative_2021,
    title = {A {Comparative} {Study} of {Existing} {Machine} {Learning} {Approaches} for {Parkinson}'s {Disease} {Detection}},
    volume = {67},
    number = {1},
    journal = {IETE Journal of Research},
    author = {Pahuja, Gunjan and Nagabhushan, T. N.},
    year = {2021},
    pages = {4--14},
}

@inproceedings{nautsch_gdpr_2019,
    title = {The {GDPR} \& {Speech} {Data}: {Reflections} of {Legal} and {Technology} {Communities}, {First} {Steps} {Towards} a {Common} {Understanding}},
    booktitle = {Interspeech},
    publisher = {ISCA},
    author = {Nautsch, Andreas and Jasserand, Catherine and Kindt, Els and Todisco, Massimiliano and Trancoso, Isabel and Evans, Nicholas},
    year = {2019},
    pages = {3695--3699},
}

@inproceedings{franzreb_towards_2024,
    title = {Towards {Audiovisual} {Anonymization} for {Remote} {Psychotherapy}: a {Subjective} {Evaluation}}, 
    booktitle = {4th {Symposium} on {Security} and {Privacy} in {Speech} {Communication}},
    publisher = {ISCA},
    author = {Franzreb, Carlos and Das, Arnab and Gieseler, Hannes and Jahn, Eva Charlotte and Polzehl, Tim and Möller, Sebastian},
    year = {2024},
    pages = {102--110},
}

@inproceedings{jin_voice_2009,
    title = {Voice convergin: {Speaker} de-identification by voice transformation},
    booktitle = {ICASSP},
    publisher = {IEEE},
    author = {Jin, Qin and Toth, Arthur R. and Schultz, Tanja and Black, Alan W.},
    year = {2009},
    pages = {3909--3912},
}

@inproceedings{meyer_use_2025,
    title = {Use {Cases} for {Voice} {Anonymization}},
    booktitle = {5th {Symposium} on {Security} and {Privacy} in {Speech} {Communication}},
    publisher = {ISCA},
    author = {Meyer, Sarina and Vu, Ngoc Thang},
    year = {2025},
}

@inproceedings{ghosh_anonymising_2024,
    title = {Anonymising {Elderly} and {Pathological} {Speech}: {Voice} {Conversion} {Using} {DDSP} and {Query}-by-{Example}},
    booktitle = {Interspeech},
    publisher = {ISCA},
    author = {Ghosh, Suhita and Jouaiti, Melanie and Das, Arnab and Sinha, Yamini and Polzehl, Tim and Siegert, Ingo and Stober, Sebastian},
    year = {2024},
    pages = {4438--4442},
}

@article{tayebi_arasteh_addressing_2024,
    title = {Addressing challenges in speaker anonymization to maintain utility while ensuring privacy of pathological speech},
    volume = {4},
    number = {1},
    journal = {Communications Medicine},
    author = {Tayebi Arasteh, Soroosh and Arias-Vergara, Tomás and Pérez-Toro, Paula Andrea and Weise, Tobias and Packhäuser, Kai and Schuster, Maria and Noeth, Elmar and Maier, Andreas and Yang, Seung Hee},
    year = {2024},
    pages = {182},
}

@inproceedings{hintz_anonymization_2023,
    title = {Anonymization of {Stuttered} {Speech} -- {Removing} {Speaker} {Information} while {Preserving} the {Utterance}},
    author = {Hintz, Jan and Bayerl, Sebastian and Sinha, Yamini and Ghosh, Suhita and Schubert, Martha and Stober, Sebastian and Riedhammer, Korbinian and Siegert, Ingo},
    year = {2023},
    pages = {41--45},
    booktitle = {3rd {Symposium} on {Security} and {Privacy} in {Speech} {Communication}},
    publisher = {ISCA},
}

@article{diaz-asper_navigating_2025,
    title = {Navigating the tradeoff between personal privacy and data utility in speech anonymization for clinical research},
    volume = {8},
    number = {1},
    journal = {npj Digital Medicine},
    author = {Diaz-Asper, Catherine and Bongo, Lars Ailo and Elvevåg, Brita},
    year = {2025},
    pages = {616},
}

@article{baevski_wav2vec_2020,
    title = {wav2vec 2.0: {A} framework for self-supervised learning of speech representations},
    volume = {33},
    journal = {Advances in Neural Information Processing Systems},
    author = {Baevski, Alexei and Zhou, Yuhao and Mohamed, Abdelrahman and Auli, Michael},
    year = {2020},
    pages = {12449--12460},
}

@inproceedings{orozco-arroyave_new_2014,
    title = {New {Spanish} speech corpus database for the analysis of people suffering from {Parkinson}'s disease},
    booktitle = {LREC},
    publisher = {European Language Resources Association (ELRA)},
    author = {Orozco-Arroyave, Juan Rafael and Arias-Londoño, Julián David and Vargas-Bonilla, Jesús Francisco and González-Rátiva, María Claudia and Nöth, Elmar},
    year = {2014},
    pages = {342--347},
}

@inproceedings{baas_voice_2023,
    title = {Voice {Conversion} {With} {Just} {Nearest} {Neighbors}},
    booktitle = {Interspeech},
    publisher = {ISCA},
    author = {Baas, Matthew and van Niekerk, Benjamin and Kamper, Herman},
    year = {2023},
    pages = {2053--2057},
}

@article{ackermann_articulatory_1991,
    title = {Articulatory deficits in parkinsonian dysarthria: an acoustic analysis.},
    volume = {54},
    number = {12},
    journal = {Journal of Neurology, Neurosurgery, and Psychiatry},
    author = {Ackermann, H and Ziegler, W},
    year = {1991},
    pages = {1093--1098},
}

@article{skodda_progression_2013,
    title = {Progression of {Voice} and {Speech} {Impairment} in the {Course} of {Parkinson}'s {Disease}: {A} {Longitudinal} {Study}},
    volume = {2013},
    journal = {Parkinson's Disease},
    author = {Skodda, S. and Grönheit, W. and Mancinelli, N. and Schlegel, U.},
    year = {2013},
    pages = {389195},
}

@article{mendes-laureano_neurovoz_2024,
    title = {{NeuroVoz}: a {Castillian} {Spanish} corpus of parkinsonian speech},
    volume = {11},
    number = {1},
    journal = {Scientific Data},
    publisher = {Nature Publishing Group},
    author = {Mendes-Laureano, Janaína and Gómez-García, Jorge A. and Guerrero-López, Alejandro and Luque-Buzo, Elisa and Arias-Londoño, Julián D. and Grandas-Pérez, Francisco J. and Godino-Llorente, Juan I.},
    year = {2024},
    pages = {1367},
}

@misc{franzreb_content_2026,
    title = {Content {Leakage} in {LibriSpeech} and {Its} {Impact} on the {Privacy} {Evaluation} of {Speaker} {Anonymization}}, 
    publisher = {arXiv},
    author = {Franzreb, Carlos and Das, Arnab and Polzehl, Tim and Möller, Sebastian},
    year = {2026},
    }

@inproceedings{gengembre_disentangling_2024,
    title = {Disentangling prosody and timbre embeddings via voice conversion},
    language = {en},
    booktitle = {Interspeech},
    publisher = {ISCA},
    author = {Gengembre, Nicolas and Le Blouch, Olivier and Gendrot, Cédric},
    year = {2024},
    pages = {2765--2769},
}

@inproceedings{guevara_2020_crowdsourcing,
    title = {{Crowdsourcing Latin American Spanish for Low-Resource Text-to-Speech}},
    author = {Guevara-Rukoz, Adriana and Demirsahin, Isin and He, Fei and Chu, Shan-Hui Cathy and Sarin, Supheakmungkol and Pipatsrisawat, Knot and Gutkin, Alexander and Butryna, Alena and Kjartansson, Oddur},
    booktitle = {LREC},
    year = {2020},
    publisher = {European Language Resources Association (ELRA)},
    pages = {6504--6513},
  }

@article{vasquez-correa_towards_2018,
    title = {Towards an automatic evaluation of the dysarthria level of patients with {Parkinson}'s disease},
    volume = {76},
    language = {en},
    journal = {Journal of Communication Disorders},
    author = {Vásquez-Correa, J.C. and Orozco-Arroyave, J.R. and Bocklet, T. and Nöth, E.},
    year = {2018},
    pages = {21--36},
}

@article{radford_robust_2022,
    title = {Robust {Speech} {Recognition} via {Large}-{Scale} {Weak} {Supervision}},
    journal = {arXiv preprint arXiv:2212.04356},
    author = {Radford, Alec and Kim, Jong Wook and Xu, Tao and Brockman, Greg and McLeavey, Christine and Sutskever, Ilya},
    year = {2022},
}

@INPROCEEDINGS{rubner1998metric,
    author={Rubner, Y. and Tomasi, C. and Guibas, L.J.},
    booktitle={Sixth International Conference on Computer Vision}, 
    publisher={IEEE},
    title={A metric for distributions with applications to image databases}, 
    year={1998},
    volume={},
    number={},
    pages={59-66},
}

@book{cover_elements_2001,
    title = {Elements of information theory},
    isbn = {978-0-471-24195-9},
    publisher = {Wiley},
    author = {Cover, Thomas M. and Thomas, Joy A.},
    year = {2001},
}

@phdthesis{champion_anonymizing_2023,
    title = {Anonymizing {Speech}: {Evaluating} and {Designing} {Speaker} {Anonymization} {Techniques}},
    school = {University of Lorraine},
    author = {Champion, Pierre},
    year = {2023},
}

@phdthesis{teixeira2024privacy,
  title={Privacy-preserving machine learning for remote speech processing},
  author={Teixeira, Francisco},
  year={2024},
  school={IST-Universidade de Lisboa, Portugal}
}

@inproceedings{teixeira2019privacy,
    title={Privacy-preserving paralinguistic tasks},
    author={Teixeira, Francisco and Abad, Alberto and Trancoso, Isabel},
    booktitle = {ICASSP},
    pages={6575--6579},
    year={2019},
    organization={IEEE}
}

@article{ravi2024enhancing,
    title = {Enhancing accuracy and privacy in speech-based depression detection through speaker disentanglement},
    journal = {Computer Speech \& Language},
    volume = {86},
    pages = {101605},
    year = {2024},
    author = {Vijay Ravi and Jinhan Wang and Jonathan Flint and Abeer Alwan},
}

@article{chen_wavlm_2022,
    title = {{WavLM}: {Large}-{Scale} {Self}-{Supervised} {Pre}-{Training} for {Full} {Stack} {Speech} {Processing}},
    volume = {16},
    number = {6},
    journal = {IEEE Journal of Selected Topics in Signal Processing},
    author = {Chen, Sanyuan and Wang, Chengyi and Chen, Zhengyang and Wu, Yu and Liu, Shujie and Chen, Zhuo and Li, Jinyu and Kanda, Naoyuki and Yoshioka, Takuya and Xiao, Xiong and Wu, Jian and Zhou, Long and Ren, Shuo and Qian, Yanmin and Qian, Yao and Wu, Jian and Zeng, Michael and Yu, Xiangzhan and Wei, Furu},
    year = {2022},
    pages = {1505--1518},
}

@inproceedings{franzreb_private_2025,
    title = {Private {kNN}-{VC}: {Interpretable} {Anonymization} of {Converted} {Speech}},
    booktitle = {Interspeech},
    publisher = {ISCA},
    author = {Franzreb, Carlos and Das, Arnab and Polzehl, Tim and Möller, Sebastian},
    year = {2025},
    pages = {3224--3228},
}

@misc{tomashenko2024voiceprivacy,
    title = {The {VoicePrivacy} 2024 {Challenge} {Evaluation} {Plan}},
    url = {http://arxiv.org/abs/2404.02677},
    doi = {10.48550/arXiv.2404.02677},
    urldate = {2025-11-04},
    publisher = {arXiv},
    author = {Tomashenko, Natalia and Miao, Xiaoxiao and Champion, Pierre and Meyer, Sarina and Wang, Xin and Vincent, Emmanuel and Panariello, Michele and Evans, Nicholas and Yamagishi, Junichi and Todisco, Massimiliano},
    month = jun,
    year = {2024},
    note = {arXiv:2404.02677 [eess]},
}

@inproceedings{franzreb_optimizing_2025,
    title = {Optimizing the {Dataset} for the {Privacy} {Evaluation} of {Speaker} {Anonymizers}},
    urldate = {2025-08-13},
    booktitle = {5th {Symposium} on {Security} and {Privacy} in {Speech} {Communication}},
    publisher = {ISCA},
    author = {Franzreb, Carlos and Das, Arnab and Polzehl, Tim and Möller, Sebastian},
    year = {2025},
}

@inproceedings{pratap20_interspeech,
    title     = {{MLS: A Large-Scale Multilingual Dataset for Speech Research}},
    author    = {Vineel Pratap and Qiantong Xu and Anuroop Sriram and Gabriel Synnaeve and Ronan Collobert},
    year      = {2020},
    booktitle = {{Interspeech}},
    pages     = {2757--2761},
    publisher = {ISCA},
}

@inproceedings{dawalatabad_ecapa-tdnn_2021,
    title = {{ECAPA}-{TDNN} {Embeddings} for {Speaker} {Diarization}},
    booktitle = {Interspeech},
    publisher = {ISCA},
    author = {Dawalatabad, Nauman and Ravanelli, Mirco and Grondin, François and Thienpondt, Jenthe and Desplanques, Brecht and Na, Hwidong},
    year = {2021},
    pages = {3560--3564},
}

@inproceedings{champion_are_2022,
    title = {Are disentangled representations all you need to build speaker anonymization systems?},
    doi = {10.21437/Interspeech.2022-10586},
    booktitle = {Interspeech 2022},
    publisher = {ISCA},
    author = {Champion, Pierre and Jouvet, Denis and Larcher, Anthony},
    year = {2022},
    pages = {2793--2797},
}

@inproceedings{kong_hifi-gan_2020,
    title = {{HiFi}-{GAN}: {Generative} {Adversarial} {Networks} for {Efficient} and {High} {Fidelity} {Speech} {Synthesis}},
    volume = {33},
    shorttitle = {{HiFi}-{GAN}},
    booktitle = {NeurIPS},
    publisher = {Curran Associates, Inc.},
    author = {Kong, Jungil and Kim, Jaehyeon and Bae, Jaekyoung},
    year = {2020},
    pages = {17022--17033},
}

@inproceedings{sicherman_analysing_2023,
    title = {Analysing {Discrete} {Self} {Supervised} {Speech} {Representation} {For} {Spoken} {Language} {Modeling}},
    booktitle = {ICASSP},
    author = {Sicherman, Amitay and Adi, Yossi},
    year = {2023},
    pages = {1--5},
    publisher={IEEE},
}

\end{document}